\documentclass[aps,prd,twocolumn,groupedaddress,nofootinbib,longbibliography]{revtex4-1}
\usepackage{amsmath,amssymb}
\usepackage{graphicx}
\usepackage{color}
\usepackage{caption}
\usepackage{hyperref}
\begin{document}


\title{AdS and dS black hole solutions in analogue gravity: \\ The relativistic and non-relativistic cases\\[10pt]}

\author{Ramit Dey}
\email[]{rdey@sissa.it}
\author{Stefano Liberati}
\email[]{stefano.liberati@sissa.it}
\author{Rodrigo Turcati}
\email[]{rturcati@sissa.it}
\affiliation{SISSA, Via Bonomea 265, 34136 Trieste, Italy
                                       and \\
     INFN sezione di Trieste, Via Valerio 2, 34127 Trieste, Italy.}

\def\d{{\mathrm{d}}}
\newcommand{\scri}{\mathscr{I}}
\newcommand{\sun}{\ensuremath{\odot}}
\def\J{{\mathscr{J}}}
\def\L{{\mathscr{L}}}
\def\sech{{\mathrm{sech}}}
\def\T{{\mathcal{T}}}
\def\tr{{\mathrm{tr}}}
\def\diag{{\mathrm{diag}}}
\def\ln{{\mathrm{ln}}}
\def\Horava{Ho\v{r}ava}
\def\Aether{\AE{}ther}
\def\AEther{\AE{}ther}
\def\aether{\ae{}ther}
\def\UH{{\text{\sc uh}}} 
\def\KH{{\text{\sc kh}}}

\begin{abstract}

\noindent We show that Schwarzschild black hole solutions in asymptotically Anti--de Sitter (AdS) and de Sitter (dS) spaces may, up to a conformal factor, be reproduced in the framework of analogue gravity. The aforementioned derivation is performed using relativistic and non-relativistic Bose--Einstein condensates. In addition, we demonstrate that the (2+1) planar AdS black hole can be mapped into the non-relativistic acoustic metric. Given that AdS black holes are extensively employed in the gauge/gravity duality, we then comment on the possibility to study the AdS/CFT correspondence and gravity/fluid duality from an analogue gravity perspective.
\end{abstract}
\maketitle


\section{Introduction}

It is well known that kinematic aspects of classical and quantum theories in curved spacetimes can be understood by using condensed matter systems which admits a hydrodynamical description \cite{Novello:2002qg,Barcelo:2005fc}. More precisely, there is a theorem which states that linearized perturbations over a moving fluid may, under suitable conditions,---barotropic, inviscid and irrotational flows---be described in the same way as massless scalar fields propagating in a curved spacetime. This class of models belongs to the broader research field of {\it Analogue Models of Gravity} {which has led to several interesting outcomes in the past decades}, such as the understanding of the robustness of Hawking radiation against high energy physics~\cite{Unruh:1994je} as well as its realisation in a laboratory setting~\cite{Brout:1995wp} or the simulation of cosmological solutions~\cite{Barcelo:2003wu}.

Amongst the various gravity systems investigated so far, black holes are the most studied ones and still subject of intense investigation both for their theoretical implications as well as for their possible realisation in laboratory.  In Refs.~\cite{Visser:1997ux,Cropp:2016teb}, for instance, the acoustic geometry have been mapped into the Schwarzschild metric. The equivalence is not exact, but up to a conformal factor, which is enough for analysing basic features of Hawking radiation. Spherically symmetric flows of incompressible fluid, which gives rise to canonical acoustic black holes, have been under investigation as well \cite{Visser:1997ux,Cropp:2016teb, Barcelo:2005fc} and even black hole solutions of gravitational theories with intrinsic preferred frames have been modelled in the analogue gravity framework \cite{Cropp:2016teb}. On the experimental side, flow configurations with analogue horizons have been reproduced in concrete laboratory set-ups leading to observations of c
 lassical and quantum features of the emission from black hole configurations~\cite{Rousseaux:2007is,Weinfurtner:2010nu,Lahav:2009wx,Steinhauer:2015saa}. Furthermore, there has been a growing interest in the attempt to simulate rotating black holes for understanding superradiance from the analogue gravity point of view as well as for testing theoretical predictions in the lab~\cite{Slatyer:2005ty,Richartz:2014lda,Cardoso:2016zvz}.

In an attempt to further extend the class of spacetimes reproducible within an analogue gravity framework, it was recently shown that one can mimic a spatial slice of an Anti--deSitter (AdS) black hole which has the same dimension as the boundary of the bulk spacetime~\cite{Hossenfelder:2014gwa,Hossenfelder:2015pza}. Obviously, the interest in this class of solutions goes beyond the mere issue of experimental realizability as they could be used as well to probe tantalising features of gravitation such as the AdS/CFT~\cite{Klebanov:2000me} and gravity/fluid duality~\cite{Hubeny:2011hd}. For example it would be interesting to understand if the latter could be seen as a natural outcome in an emergent gravity framework as that epitomised by analogue gravity. For example, in Refs.~\cite{Hossenfelder:2014gwa,Hossenfelder:2015pza} it was conjectured---inside the context of gauge/gravity duality---that there may exist a connection between the weakly coupled condensed matter system mimicking 
 a projection of the bulk and the strongly coupled condensed matter living at the boundary. In addition, attempts to investigate the AdS/CFT correspondence in the formalism of analogue gravity have also been explored in Refs.~\cite{Das:2010mk,Semenoff:2012xu,Chen:2012uc,Khveshchenko:2013foa,Bilic:2014dda,Ge:2015uaa} albeit in these cases the analogue system was built up at the boundary, and not in the bulk, as in Refs.~\cite{Hossenfelder:2014gwa,Hossenfelder:2015pza}. 

Unfortunately, an inherent obstruction for drawing stronger conclusions about these tantalising features of gravitational theories in AdS spacetimes is given by the fact that the Einstein equations are not generically reproduced by the background field equations of the analogue gravity system.  Nonetheless, simpler gravitational theories can be reproduced within the analogue gravity framework where Bose--Einstein condensates were used to reproduce Newtonian~\cite{Girelli:2008gc} as well as N\"ordstrom scalar gravity~\cite{Belenchia:2014hga} with a cosmological constant. Since in the above mentioned scenario we have a well defined fluid, which gives rise to a gravity theory in the bulk, one could explore if there is some deeper connection between the fluid living in the bulk and the conformal theory at the boundary (which admits a fluid description leading to the so called gravity/fluid duality). 

Our aim in this paper is then twofold. 

\begin{enumerate}
 \item To show that black holes in asymptotic AdS and dS spaces can be formulated in the context of analogue gravity by using relativistic and non-relativistic moving fluids (without need for projections to lower dimensions). 
 
\item To provide an insight towards a possible use of analogue gravity as a toy model for understanding features of the gauge-fluid/gravity duality in a scenario of emergent gravity dynamics. 
 
\end{enumerate}

The remainder of this paper is organized as follows. In Sec.~\ref{uncoupledrbec} we review the basic ingredients of the Bose--Einstein condensate formation in the relativistic regime, showing the emergence of the effective acoustic metric for the linearized perturbations. The non-relativistic regime is obtained afterwards. In Sec.~\ref{AdSdSbhsection} we derive the Schwarzschild AdS and dS acoustic black hole metrics in arbitrary dimensions using the relativistic condensate. In Sec.~\ref{AdSdSbhnonrelativisticsection} we derive the AdS and dS black hole for the non-relativistic acoustic metric. In Sec.~\ref{planarAdSbhsection}, we map the (2+1)D planar AdS black hole into the non-relativistic acoustic metric. At the end, in Sec.~\ref{SummaryandDiscussion}, we present our conclusions and make a discussion about the gauge/gravity duality in the context of emergent gravity dynamics from the analogue gravity perspective.

In our conventions the signature of the metric is $(-,+,...,+)$.

\section{Bose--Einstein condensates}\label{uncoupledrbec}

In this section we review the basic features of the Bose--Einstein condensate in both relativistic and non-relativistic regime and how the quasiparticles over them can be described in terms of an effective acoustic metric. For further details, we refer to Ref.~\cite{Fagnocchi:2010sn}. Let us start by considering the Lagrangian density for a complex scalar field $\phi(\mathbf{x},t)$, which can be written as 
\begin{eqnarray}\label{lrbec}
\mathcal{L}=-\eta^{\mu\nu}\partial_{\mu}\phi^{*}\partial_{\nu}\phi-\left(\frac{m^{2}c^{2}}{\hbar^{2}}+V(t,\mathbf{x})\right)\phi^{*}\phi
\nonumber\\
-U(\phi^{*}\phi;\lambda_{i}),
\end{eqnarray}
where $m$ is the mass of bosons, $V(t,\mathbf{x})$ is an external potential, $c$ is the speed of light, $U$ is a self-interaction term and $\lambda_{i}(t,\mathbf{x})$ are the coupling constants. 

The Lagrangian (\ref{lrbec}) is invariant under the global $U(1)$ symmetry and has a conserved current 
\begin{equation}
j^{\mu}=i(\phi^{*}\partial^{\mu}\phi-\phi\partial^{\mu}\phi^{*}), 
\end{equation}
which is related to a conserved ensemble charge $N-\bar{N}$, where $N(\bar{N})$ is the number of bosons (anti-bosons).

In the absence of self-interactions ($U=0$) and no external potential ($V=0$), the average number of bosons $n_{k}$ in a state of energy $E_{k}$ can be written as
\begin{equation}\label{nb}
N-\bar{N}=\Sigma_{k}[n_{k}-\bar{n}_{k}], 
\end{equation}
where 
\begin{eqnarray}
n_{k}(\mu,\beta)=1/\left\lbrace{\exp}[\beta(|E_{k}|-\mu)]-1\right\rbrace, 
\nonumber\\
  \bar{n}_{k}(\mu,\beta)=1/\left\lbrace{\exp[\beta(|E_{k}|+\mu)]-1}\right\rbrace,
\end{eqnarray}
 $\mu$ is the chemical potential, $T\equiv1/(k_{B}\beta)$ is the temperature and the energy of the state $k$ is given by $E^{2}_{k}=\hbar^{2}k^{2}c^{2}+m^{2}c^{4}$.

In a system of volume $\mathcal{V}$, the relation between the conserved charge density, $n=(N-\bar{N})/\mathcal{V}$, and the critical temperature $T_c$ is
\begin{equation}\label{nd}
n=C\int^{\infty}_{0}\d kk^{d-1}\frac{\sinh(\beta_{c}mc^{2})}{\cosh(\beta_{c}|E_{k}|)-\cosh(\beta_{c}mc^{2})}, 
\end{equation}
where $C=1/(2^{d-1}\pi^{d/2}\Gamma(d/2))$ and $\beta_c\equiv1/(k_{B}T_c)$.

The non-relativistic and ultra-relativistic limits can be obtained directly from Eq.~(\ref{nd}). When $k_{B}T_{c}\ll{mc^{2}}$, the non-relativistic limit is given by
\begin{equation}
k_{B}T_{c}=\frac{2\pi\hbar^{2}}{n}\left(\frac{n}{\zeta(d/2)}\right)^{2/d}, 
\end{equation}
where $\zeta$ is the Riemann zeta function. The ultra-relativistic limit is characterized by the condition $k_{B}T_{c}\gg{mc^{2}}$, which implies that
\begin{equation}
(k_{B}T_{c})^{d-1}=\frac{\hbar^{d}c^{d-2}\Gamma(d/2)(2\pi)^{d}}{4m\pi^{d/2}\Gamma(d)\zeta(d-1)}n.
\end{equation}

The condensation of the relativistic Bose gas occurs when $T\ll{T_{c}}$, where, using the mean-field approximation, the dynamics of the condensate is described by the nonlinear Klein--Gordon equation
\begin{equation}\label{groundstateequation}
\Box\phi-\left(\frac{m^{2}c^{2}}{\hbar^{2}}+V\right)\phi-U'\phi=0. 
\end{equation}

In this phase, it is possible to uncouple the BEC ground state from its perturbations. To perform this split one can insert $\phi=\varphi(1+\psi)$ in Eq.~(\ref{groundstateequation}), where $\varphi$ is the classical background field satisfying the equation
\begin{equation}\label{nlkg}
\Box\varphi-\left(\frac{m^{2}c^{2}}{\hbar^{2}}+V\right)\varphi-U'\varphi=0, 
\end{equation}
and $\psi$ is a quantum relative fluctuation (i.e., of order $\hbar$). The modified Klein-Gordon equation (\ref{nlkg}) gives the dynamics of the ground state of the relativistic condensate.

It is also convenient to decompose the degrees of freedom of the complex classical scalar field in terms of the Madelung representation, $\varphi=\sqrt{\rho}e^{i\theta}$. Using this prescription, the continuity equation and the condensate equation (\ref{nlkg}) assume the form
\begin{eqnarray}\label{cemrnc}
\partial_{\mu}(\rho{u^{\mu}})&=&0, \\
-u_{\mu}u^{\mu}&=&c^{2}+\frac{\hbar^{2}}{m^{2}}\left[V(x^{\mu})+U'(\rho;\lambda_{i}(x^{\mu}))-\frac{\Box\sqrt{\rho}}{\rho}\right],\nonumber\\
\end{eqnarray}
where 
\begin{equation}\label{hypersurface}
u^{\mu}=\frac{\hbar}{m}\partial^{\mu}\theta
\end{equation}
is the unnormalized four-velocity of the condensate.

The linearized perturbation $\psi$ satisfies 
\begin{eqnarray}\label{lpe}
\left\lbrace\left[i\hbar{{u}^{\mu}}\partial_{\mu}+T_{\rho}\right]\frac{1}{c_{0}^{2}}\left[-i\hbar{{u}^{\nu}}\partial_{\nu}+T_{\rho}\right]-\frac{\hbar^{2}}{\rho}\eta^{\mu\nu}\partial_{\mu}\rho\partial_{\nu}\right\rbrace\psi=0,\nonumber\\
\end{eqnarray}
where $c_{0}^{2}\equiv\frac{\hbar^{2}}{2m^{2}}\rho{U''}$ is related to the interaction strength and 
\begin{equation}
T_{\rho}\equiv-\frac{\hbar^{2}}{2m}(\Box+\eta^{\mu\nu}\partial_{\mu} \mathrm{ln} \rho\partial_{\nu})
\end{equation}
is a generalized kinetic operator. Albeit $c_{0}$ has dimension of velocity, it is not the speed of sound in the relativistic condensate. Nevertheless, as we will see in the relation (\ref{speedofsound}), both variables are connected.

Equation (\ref{lpe}) is the relativistic generalization of the Bogoliubov--de Gennes equation. The dispersion relation associated to~\eqref{lpe} has several limiting cases of interest which were fully explored in Ref.~\cite{Fagnocchi:2010sn}. Here, we are particularly interested in the low momentum regime which is characterized by the condition
\begin{equation}\label{analoguecondition}
|k|\ll\frac{mu^{0}}{\hbar}\left(1+b\right).
\end{equation}
where $b\equiv(c_{0}/u^{0})^{2}$. Taking into account the phononic limit defined by (\ref{analoguecondition}) and assuming that the background quantities $u$, $\rho$ and $c_{0}$ vary slowly in space and time on scales comparable with the wavelength of the perturbation, i.e.,
\begin{eqnarray}
\left|\frac{\partial_{t}\rho}{\rho}\right|\ll{w}, \quad\quad \left|\frac{\partial_{t}c_{0}}{c_{0}}\right|\ll{w}, \quad\quad \left|\frac{\partial_{t}u_{\mu}}{u_{\mu}}\right|\ll{w}, 
\end{eqnarray}
one can disregard the quantum potential $T_{\rho}$, implying that the quasiparticles can be described in terms of an effective acoustic metric. Let us then show that the acoustic description can be achieved at the aforementioned scales. Applying the above assumptions, it is easy to see that Eq.~(\ref{lpe}) reduces to
\begin{eqnarray}\label{lpnqp}
\left[{{u}^{\mu}}\partial_{\mu}\left(\frac{1}{c_{0}^{2}}{{u}^{\nu}}\partial_{\nu}\right)-\frac{1}{\rho}\eta^{\mu\nu}\partial_{\mu}\left(\rho\partial_{\nu}\right)\right]\psi=0.
\end{eqnarray}

Now, in order to arrive at the acoustic metric, one can make use of the continuity equation (\ref{cemrnc}) and rewrite (\ref{lpnqp}) as
\begin{eqnarray}\label{phononicequation}
\partial_{\mu}\left[-\rho\eta^{\mu\nu}+\frac{\rho}{c_{0}^{2}}u^{\mu}u^{\nu}\right]\partial_{\nu}\psi=0.
\end{eqnarray}

Equation (\ref{phononicequation}) can be expressed as
\begin{equation}
\partial_{\mu}\left(\gamma^{\mu\nu}\partial_{\nu}{\psi}\right)=0, 
\end{equation}
where $\gamma^{\mu\nu}$ is 
\begin{equation}
\gamma^{\mu\nu}=\frac{\rho}{c_{0}^{2}}\left(c_{0}^{2}\eta^{\mu\nu}-u^{\mu}u^{\nu}\right).
\end{equation}

Identifying $\gamma^{\mu\nu}=\sqrt{-g}g^{\mu\nu}$, we get
\begin{equation}
\sqrt{-g}=\rho^{2}\sqrt{1-u^{\alpha}u_{\alpha}/c_{0}^{2}}, 
\end{equation}
and
\begin{eqnarray}
g^{\mu\nu}=\frac{1}{\rho{c}_{0}^{2}\sqrt{1-u^{\alpha}u_{\alpha}/c_{0}^{2}}}\left(c_{0}^{2}\eta^{\mu\nu}-u^{\mu}u^{\nu}\right).
\end{eqnarray}

Therefore, one can cast Eq.~(\ref{phononicequation}) in the form  
\begin{equation}\label{dalembertian}
\triangle{\psi}\equiv\frac{1}{\sqrt{-g}}\partial_{\mu}\left(\sqrt{-g}g^{\mu\nu}\partial_{\nu}{\psi}\right),
\end{equation}
which is a D'Alembertian for a massless scalar in a curved background. Inverting $g^{\mu\nu}$, one can then see that the acoustic metric $g_{\mu\nu}$ for the quasiparticles propagation in a (3+1)D relativistic, barotropic, irrotational fluid flow is given by
\begin{eqnarray}\label{ramnr}
g_{\mu\nu}=\frac{\rho}{\sqrt{1-u_{\alpha}u^{\alpha}/c_{0}^{2}}}\left[\eta_{\mu\nu}\left(1-\frac{u_{\alpha}u^{\alpha}}{c_{0}^{2}}\right)+\frac{u_{\mu}u_{\nu}}{c_{0}^{2}}\right]. 
\end{eqnarray}

Sometimes is more convenient express the acoustic metric (\ref{ramnr}) as
\begin{equation}\label{ram}
g_{\mu\nu}=\left(\rho\frac{{c}}{c_{s}}\right)^{2/d-1}\left[\eta_{\mu\nu}+\left(1-\frac{c_{s}^{2}}{c^{2}}\right)\frac{v_{\mu}v_{\nu}}{c^{2}}\right],
\end{equation}
where 
\begin{equation}\label{normalizedfourvelocity}
v^{\mu}=c\frac{u^{\mu}}{||u||} 
\end{equation}
is the normalized four-velocity and the speed of sound $c_{s}$ is defined by
\begin{equation}\label{speedofsound}
c_{s}^{2}=\frac{c^{2}c_{0}^{2}/||u||^{2}}{1+c_{0}^{2}/||u||^{2}}.
\end{equation}

It is obvious from Eq.~(\ref{ram}) that the acoustic metric $g_{\mu\nu}$ is related via a disformal transformation to the background Minkowski spacetime. Writing in the lab coordinates ($x^{\mu}\equiv{c}t,x^{i}$), the relativistic acoustic line element takes the form
\begin{eqnarray}
ds^{2}=\left(\rho\frac{{c}}{c_{s}}\right)^{\frac{2}{d-1}}\left[\left(-1+\xi\frac{{v_0}^{2}}{c^{2}}\right)c^{2}dt^{2}+2\xi\frac{v_{0}v_{i}}{c^{2}}cdtdx^{i}+\right.
\nonumber\\
\left. \left(\delta_{ij}+\xi\frac{v_{i}v_{j}}{c^{2}}\right)dx^{i}dx^{j}\right],\nonumber\\
\end{eqnarray}
where $\xi\equiv\left(1-c_{s}^{2}/c^{2}\right)$. The normalization condition $v^{2}=-c^{2}$ allow us to rewrite the above acoustic element line as 
\begin{eqnarray}\label{relacoustmetric}
ds^{2}=\left(\rho\frac{{c}}{c_{s}}\right)^{\frac{2}{d-1}}\left[-\left(c_{s}^{2}-\xi\mathbf{v}^{2}\right)dt^{2}\pm2\xi\sqrt{1+\frac{\mathbf{v}^{2}}{c^{2}}}(v_{i}dx^{i})dt
\right.
\nonumber\\
\left.
+\left(\delta_{ij}+\xi\frac{v_{i}v_{j}}{c^{2}}\right)dx^{i}dx^{j}\right],\nonumber\\
\end{eqnarray}
where $\mathbf{v}^{2}=v_{i}v^{i}$ is the square normalized 3-velocity. 

To find the non-relativistic limit of the above acoustic metric, one needs to make some further assumptions. First, in the non-relativistic regime the speed of sound $c_{s}$ and the flow velocity $v^{i}$ are much smaller than the speed of light, which implies that
\begin{eqnarray}
\xi=1+\frac{c_{s}^{2}}{c^{2}}\approx1, \quad\quad\quad 1+\frac{\mathbf{v}^{2}}{c^{2}}\approx1.
\end{eqnarray}

Moreover, the self-interaction between the atoms must be weak, i.e. $c_{0}\ll{c}$. Besides that, $u^{0}\rightarrow{c}$. In addition, we note that the speed of sound (\ref{speedofsound}) goes to $c_{0}$. Under these assumptions, the four-velocity takes the form $v^{\mu}\approx(c;u^{i})$. Finally, applying all these condition into the relativistic acoustic metric (\ref{relacoustmetric}), we promptly arrive at
\begin{eqnarray}\label{nonrelativisticacousticmetric}
ds^{2}=\frac{\rho_{m}}{c_{s}}\left[-\left(c_{s}^{2}-{\mathbf{u}^{2}}\right)dt^{2}\pm2u_{i}dx^{i}dt+\delta_{ij}dx^{i}dx^{j}\right],
\end{eqnarray}
where $\rho_{m}$ is the mass density and $\mathbf{u}^{2}=u_{i}u^{i}$ is the square unnormalized 3-velocity. The form of the line element (\ref{nonrelativisticacousticmetric}) is the usual non-relativistic acoustic metric. 


\section{Analogue of Schwarzschild--AdS and dS black holes\\ in relativistic BEC}\label{AdSdSbhsection}

We now turn our attention to black holes in asymptotic AdS and dS spaces. The line element of these geometries are given by
\begin{eqnarray}\label{AdSdSBH}
ds^{2}=-\left(1-\frac{r_{0}}{r^{d-2}}\pm\frac{r^{2}}{L^{2}}\right)c_{s}^{2}d\tau^{2}
\nonumber\\
+\left(1-\frac{r_{0}}{r^{d-2}}\pm\frac{r^{2}}{L^{2}}\right)^{-1}dr^{2}+r^{2}d\Omega^{2}, 
\end{eqnarray}
where ${1}/{L^{2}}$ equals the cosmological constant $\Lambda$, $d$ is the spatial dimension, $r_0$ is a constant and the $+$ and $-$ signs pertain respectively to the AdS  and dS solutions.

Here, we will derive both these solutions in the framework of relativistic acoustic geometries. To find such a map between the relativistic flow and the aforementioned metrics, we start by rewriting (\ref{relacoustmetric}) in spherical coordinates and assuming an isotropic fluid, i.e., $v^{\theta}=v^{\phi}=0$. Considering the normalized velocity profile given by
\begin{eqnarray}\label{radialvelocity}
v_{r}^{2}&=&\frac{c_{s}^{2}}{\xi}\left({\frac{r_{0}}{r^{d-2}}\mp\frac{r^{2}}{L^{2}}}\right),
\end{eqnarray}
the relativistic acoustic metric (\ref{relacoustmetric}) assumes the form  
\begin{widetext}
\begin{eqnarray}\label{relativisticbh}
ds^{2}&=&\left(\rho\frac{{c}}{c_{s}}\right)^{\frac{2}{d-1}}\left\lbrace-\left(1-\frac{r_{0}}{r^{d-2}}\pm\frac{r^{2}}{L^{2}}\right)c_{s}^{2}dt^{2}-2\left(\sqrt{\left({\frac{r_{0}}{r^{d-2}}\mp\frac{r^{2}}{L^{2}}}\right)} \sqrt{1-\frac{c_{s}^{2}}{c^{2}}\left({1-\frac{r_{0}}{r^{d-2}}\pm\frac{r^{2}}{L^{2}}}\right)}\right)c_{s}dtdr\right.\nonumber\\
&&\left.+\left[1+\frac{c_{s}^{2}}{c^{2}}\left({\frac{r_{0}}{r^{d-2}}\mp\frac{r^{2}}{L^{2}}}\right)\right]dr^{2}+r^{2}d\Omega^{2}\right\rbrace, 
\end{eqnarray}
\end{widetext}
where the physical meaning of the $``\mp"$ signs will become clear at the end. The normalized radial velocity (\ref{radialvelocity}) implies that the unnormalized radial flow is
\begin{equation}\label{unradialvelocity}
u^{r}=\frac{u^{0}(c_{s}/\xi)\sqrt{(r_{0}/r^{d-2})\mp(r/L)^{2}}}{c\sqrt{1+(c_{s}^{2}/\xi)((r_{0}/r^{d-2})\mp(r/L)^{2})}}. 
\end{equation}

According to $(d+1)$ continuity equation (\ref{cemrnc}), it reduces to
\begin{equation}\label{ssce}
\partial_{r}\left(r^{d-1}\rho{u^{r}}\right)=0, 
\end{equation} 
which can be satisfied only for a incompressible fluid, where the density profile  $\rho$ takes the form
\begin{equation}
\rho=\left(\frac{A}{r^{d-1}}\right)\frac{c\sqrt{1+(c_{s}^{2}/\xi)((r_{0}/r^{d-2})\mp(r/L)^{2})}}{u^{0}(c_{s}/\xi)\sqrt{(r_{0}/r^{d-2})\mp(r/L)^{2}}}. 
\end{equation}
In the above relation, $A$ and the speed of sound $c_{s}$ are both position-independent factors.

Next we define a new time coordinate $\tau$ by 
\begin{widetext}
\begin{eqnarray}\label{coordtransf}
c_{s}d\tau=c_{s}dt
+\sqrt{\left({\frac{r_{0}}{r^{d-2}}\mp\frac{r^{2}}{L^{2}}}\right)}\sqrt{1-\frac{c_{s}^{2}}{c^{2}}\left({1-\frac{r_{0}}{r^{d-2}}\pm\frac{r^{2}}{L^{2}}}\right)}\left(1-\frac{r_{0}}{r^{d-2}}\pm\frac{r^{2}}{L^{2}}\right)^{-1}dr.
\end{eqnarray}
\end{widetext}
Substituting back the coordinate transformation (\ref{coordtransf}) into the relativistic acoustic line (\ref{relativisticbh}), we promptly find

\begin{eqnarray}\label{AdSdSacousticBH}
ds^{2}=\left(\rho\frac{{c}}{c_{s}}\right)^{\frac{2}{d-1}}\Bigg[-\left(1-\frac{r_{0}}{r^{d-2}}\pm\frac{r^{2}}{L^{2}}\right)c_{s}^{2}d\tau^{2}
\nonumber\\
+\left(1-\frac{r_{0}}{r^{d-2}}\pm\frac{r^{2}}{L^{2}}\right)^{-1}dr^{2}+r^{2}d\Omega^{2}\Bigg], 
\end{eqnarray}

which is conformal to the AdS and dS black hole geometries \eqref{AdSdSBH} with $L^{-2}=\Lambda$. Also, by confronting \eqref{AdSdSBH} with the above metric, one can now see that the ``-" sign in ~\eqref{relativisticbh} refers to the AdS black hole solution, while the ``+" is related to the dS black hole geometry. So in conclusion, we have found that it is possible to mimic in any number of dimensions a spacetime conformal to that  of a Schwarzschild--AdS/dS black hole with an analogue model which reduces to a relativistic fluid in some limit (just like the relativistic BEC above described).~\footnote{It must be noted that, while the fluid which define the dS acoustic black hole is well behaved everywhere in the acoustic spacetime, the AdS black hole solution is valid only in the regime in which the velocity profile (\ref{radialvelocity}) is non-zero, i.e., when the condition 
$r^{d}<r_{0}L^{2}$
is satisfied.}

\section{Analogue of AdS and dS black holes\\ in non-relativistic BEC}

\subsection{Acoustic Schwarzschild--AdS and dS black holes}\label{AdSdSbhnonrelativisticsection} 

We have so far considered only relativistic condensates in order to derive the AdS and dS acoustic black hole metrics. Now we will show that the previous solutions can also be found through the use of non-relativistic BECs. The derivation of such outcome is quite similar to the relativistic case. So, starting from the non-relativistic acoustic metric (\ref{nonrelativisticacousticmetric}), and considering a fluid only in the  $r$-direction given by
\begin{eqnarray}
u_{r}=c_{s}\sqrt{\left(\frac{r_{0}}{r^{d-2}}-\frac{r^{2}}{L^{2}}\right)}, 
\end{eqnarray}
one easily finds 
\begin{widetext}
\begin{eqnarray}\label{nonrelativisticbh}
ds^{2}=\left(\frac{\rho}{c_{s}}\right)^{\frac{2}{d-1}}\left[-\left(1-\frac{r_{0}}{r^{d-2}}+\frac{r^{2}}{L^{2}}\right)c_{s}^{2}dt^{2}-2\sqrt{\left(\frac{r_{0}}{r^{d-2}}-\frac{r^{2}}{L^{2}}\right)}dr\left(c_{s}dt\right)+dr^{2}+r^{2}d\Omega^{2}\right].\nonumber\\
\end{eqnarray}
\end{widetext}
Again, as in the relativistic case, to satisfy the continuity equation (\ref{ssce}) one needs to consider a incompressible fluid, where the density $\rho$ now is given by
\begin{equation}
\rho=\frac{B}{c_{s}}\left(\sqrt{\left(\frac{r_{0}}{r^{d-2}}-\frac{r^{2}}{L^{2}}\right)}\right)^{-1}, 
\end{equation}
and $B$ is not a function of spacetime. It means that similar to the relativistic case, at most, one can simulate the AdS and dS black hole acoustic metrics up to a conformal factor. Therefore, applying the coordinate transformation
\begin{eqnarray}
c_{s}d\tau&=&c_{s}dt+\sqrt{\left({\frac{r_{0}}{r^{d-2}}\mp\frac{r^{2}}{L^{2}}}\right)}\left(1-\frac{r_{0}}{r^{d-2}}\pm\frac{r^{2}}{L^{2}}\right)^{-1}dr \nonumber\\
\end{eqnarray}
in (\ref{nonrelativisticbh}), one gets the AdS and dS black hole metrics (\ref{AdSdSacousticBH}) up to a conformal factor. Again, as in the relativistic case, the AdS solution is valid only when the condition $r^{d}<r_{0}L^{2}$ is satisfied.


\subsection{Acoustic Planar AdS black holes}\label{planarAdSbhsection}

We have just seen that Schwarzschild black holes in asymptotic AdS and dS spacetimes can be mimicked using relativistic and non-relativistic fluids. Now, we would like to go one step further and see if it is possible to map the acoustic metric into the planar AdS black hole. This class of black holes are the most commonly used in the formalism of the gauge/gravity duality. A first attempt to obtain this class of solutions was made in Ref.~\cite{Hossenfelder:2014gwa}. The author found that such a map is feasible only when considering a projection of the bulk solution. Here we will show that, under some appropriate choices, it is possible to simulate all the bulk without any projection. 

To begin with, we note that the $(d+1)$ planar AdS black hole metric in Poincar\`e coordinates takes the form
\begin{eqnarray}\label{planarAdSbh}
ds^{2}=\frac{L^{2}}{z^{2}}\Bigg[\left(1-\frac{z^{d}}{z_{0}^{d}}\right)c^{2}dt^{2}+\left(1-\frac{z^{d}}{z_{0}^{d}}\right)^{-1}dz^{2}
\nonumber\\
+\sum_{i=1}^{d-1}dx^{i}dx^{i}\Bigg].
\end{eqnarray}

Now, considering the non-relativistic fluid acoustic metric (\ref{nonrelativisticacousticmetric}) with planar symmetry, a fluid flow only in the $z-$direction and perfoming the coordinate transformation
\begin{eqnarray}
d\tau&=&dt+\frac{u_{z}dz}{\left(c_{s}^{2}-u_{z}^{2}\right)}, 
\end{eqnarray}
we arrive at the diagonal form of the acoustic line element given by 
\begin{eqnarray}\label{labplanarbh}
ds^{2}=\left(\frac{\rho}{c_{s}}\right)^{\frac{2}{(d-1)}}\Bigg[-\left(1-\frac{u_{z}^{2}}{c_{s}^{2}}\right)c_{s}^{2}d\tau^{2}
\nonumber\\
+\left(1-\frac{u_{z}^{2}}{c_{s}^{2}}\right)^{-1}dz^{2}
+\sum_{i=1}^{d-1}dx^{i}dx^{i}\Bigg].
\end{eqnarray}

Now, comparing (\ref{planarAdSbh}) and (\ref{labplanarbh}), it is natural to identify
\begin{equation}
\left(\frac{\rho}{c_{s}}\right)^{\frac{2}{(d-1)}}=\left(\frac{L}{z}\right)^{2} \rightarrow \rho=c_{s}\left(\frac{L}{z}\right)^{d-1},
\end{equation}
and
\begin{equation}
\frac{u_{z}^{2}}{c_{s}^{2}}=\frac{z^{d}}{z_{0}^{d}} \rightarrow u_{z}=c_{s}\left(\frac{z}{z_{0}}\right)^{d/2},  
\end{equation}
where we are picking the speed of sound $c_{s}$ as position-independent.

From the above relations, the density goes to $\rho\sim{z^{1-d}}$, while the flow velocity goes to $u_{z}\sim{z^{d/2}}$. Under the planar symmetry, and taking into account that we are in the static regime, the continuity equation \eqref{cemrnc} reduces to
\begin{eqnarray}\label{ceplanar}
\partial_{z}(\rho{u^{z}})=0,
\end{eqnarray}
which can be satisfied only for $d=2$ due the relation $\rho{u_{z}}\sim {z}^{1-d/2}$. 
Hence we proved that non-relativistic acoustic metrics can {\em exactly} mimic (2+1) planar AdS black holes in the bulk.

In Ref.~\cite{Hossenfelder:2014gwa} the number of spatial dimensions of the analogue system was taken to be different from the number of dimensions of the AdS black hole that had to be mimicked. For the choice of coordinate system used in Ref.~\cite{Hossenfelder:2014gwa} and given that the analogue system is conjectured to live in (3+1) dimensions, imposing the continuity equation~\eqref{ceplanar} fixes the spatial dimensions of the planar black hole to be $d=4$. This in turn implied that only a projection of the (4+1) AdS black hole geometry to one lesser dimension could be reproduced. In our case the dimensions of the analogue system and those of the AdS black hole geometry are taken to coincide a priori which in turns implied that both must be equal to two.

However, let us stress that  the above constraints on the dimensions can be overcomed if one is interested in mimicking metrics which are just conformally related to the planar AdS black hole ~\eqref{planarAdSbh}. Indeed, in this case it is easy to check that one can always choose a conformal factor such that  $\rho{u_{z}}=\mbox{constant}$ and thus satisfying the continuity equation~\eqref{ceplanar}. 

Coming back to the possibility to {\em exactly} simulate planar black holes in arbitrary dimensions of course a natural question would be if the adoption of an analogue system based on a relativistic fluid (such as a relativistic BEC) could improve things. Unfortunately, the answer is no. 
%
To see it explicitly, we first consider a $d$-dimensional relativistic fluid flow in the $z$-direction given by
\begin{equation}\label{normalizedvelocityplanarbh}
v_{z}=\frac{c_{s}}{\xi^{1/2}}\left(\frac{z}{z_{0}}\right)^{d/2}.
\end{equation}

Now we shall bring the relativistic acoustic metric~\eqref{relacoustmetric} into the diagonal form through the coordinate transformation
\begin{equation}
c_{s}d\tau=c_{s}dt+\frac{z^{d/2}}{z_{0}^{d/2}}\sqrt{1-\frac{c_{s}^{2}}{c^{2}}\left(1-\frac{z^{d}}{z_{0}^{d}}\right)}\left(1-\frac{z^{d}}{z_{0}^{d}}\right)^{-1}dz,
\end{equation}
where, after the above transformation, Eq.~\eqref{relacoustmetric} becomes
\begin{eqnarray}\label{labplanarbh}
ds^{2}=\left(\rho\frac{c}{c_{s}}\right)^{\frac{2}{(d-1)}}\Bigg[-\left(1-\xi\frac{v_{z}^{2}}{c_{s}^{2}}\right)c_{s}^{2}d\tau^{2}
\nonumber\\
+\left(1-\xi\frac{v_{z}^{2}}{c_{s}^{2}}\right)^{-1}dz^{2}
+\sum_{i=1}^{d-1}dx^{i}dx^{i}\Bigg].
\end{eqnarray}

Comparing Eq.~\eqref{labplanarbh} and Eq.~\eqref{planarAdSbh}, we promptly find that
\begin{equation}\label{relativisticdensity}
\left(\rho\frac{c}{c_{s}}\right)^{\frac{2}{(d-1)}}=\left(\frac{L}{z}\right)^{2} \rightarrow \rho=\frac{c_{s}}{c}\left(\frac{L}{z}\right)^{d-1}.
\end{equation}

However, according to the four-normalized velocity relation (\ref{normalizedvelocityplanarbh}), the unnormalized flow velocity $u^{z}$ takes the form
\begin{equation}\label{unnormalizedvelocityprofile}
u^{z}=\frac{u^{0}}{c}\frac{c_{s}(z/z_{0})^{d/2}}{\sqrt{1+(c_{s}/c)^{2}(z/z_{0})^{d}}} 
\end{equation}
which, due to our previous choice  for the density profile Eq.~(\ref{relativisticdensity}), does not satisfy the continuity equation (\ref{cemrnc}) and even allowing a position dependence in the speed of sound $c_{s}$, one still cannot satisfy the continuity equation (\ref{ceplanar}). Again, it is worth stressing that also in this case the introduction of a conformal factor, relating the acoustic and AdS planar black hole metrics, can be used to mimic the latter while allowing at the same time to satisfy the continuity equation~\eqref{cemrnc}. To see how we can find such a conformal mapping, one can multiply Eq.~\eqref{planarAdSbh} by an arbitrary function $f(z)$, where the planar AdS black hole line element becomes
\begin{eqnarray}
ds^{2}=f(z)\frac{L^{2}}{z^{2}}\Bigg[\left(1-\frac{z^{d}}{z_{0}^{d}}\right)c^{2}dt^{2}+\left(1-\frac{z^{d}}{z_{0}^{d}}\right)^{-1}dz^{2}
\nonumber\\
+\sum_{i=1}^{d-1}dx^{i}dx^{i}\Bigg]. 
\end{eqnarray}

Assuming the same normalized velocity than in Eq.~\eqref{normalizedvelocityplanarbh}, which automatically gives the unnormalized fluid flow \eqref{unnormalizedvelocityprofile}, and picking up a position-independent speed of sound $c_{s}$,  we promptly obtain to the density profile
\begin{equation}
\left(\rho\frac{c}{c_{s}}\right)^{\frac{2}{(d-1)}}=f(z)\left(\frac{L}{z}\right)^{2} \rightarrow \rho=f(z)^{\frac{(d-1)}{2}}\left(\frac{c_{s}}{c}\right)\left(\frac{L}{z}\right)^{d-1}.
\end{equation}

It is now obvious that an appropriate choice of the function $f(z)$ will satisfy the continuity equation~\eqref{cemrnc} (in this case in any dimensions), allowing the planar AdS black hole metric be conformally mapped into the relativistic acoustic geometry.

\section{Summary and Discussion}\label{SummaryandDiscussion}

 Mapping black hole metrics is one of the most interesting applications of analogue models of gravity. However, only Schwarzschild metric and spatial sections of rotating spacetimes have been obtained so far. In the present work we have applied for the first time the analogy between Schwarzschild AdS/dS black holes and condensed matter systems using a BEC. Starting from both relativistic and non-relativistic Bose--Einsten condensates, we have conformally mapped the acoustic metric in the aforementioned solutions for BECs of arbitrary dimension. It is an interesting fact that the (2+1) dimensional planar AdS black holes can be exactly mimicked using a non-relativistic BEC. Nevertheless, higher dimensional ones can be mimicked modulo a conformal factor.




Nowadays there are several laboratory set up underway based on theoretical models that have been proposed over the past years in the analogue gravity framework. Currently, such experimental realisations include mainly non-relativistic Bose--Einstein condensates, optical fibres and shallow wave systems \cite{Barcelo:2005fc}. Albeit relativistic Bose--Einstein condensates provide us with a very rich framework to explore analogue models of gravity, differently from the non-relativistic case where some features can be simulated, the experimental realisation seems far from being achieved in a nearby future. Indeed, the use of relativistic BECs for theoretical studies is mainly focused in issues related to Hawking radiation and spacetime and gravity emergent phenomena.~\footnote{However, it is perhaps worth stressing that our results apply to general non-relativist and relativistic superfluids as well.}

In this latter declination, given that we now know how to mimic AdS black holes in analogue gravity, it would be interesting to investigate if one can acquire some knowledge about AdS/CFT~\cite{Klebanov:2000me} correspondence and gravity/fluid duality~\cite{Hubeny:2011hd} from this perspective. However, in the context of analogue gravity, the Einstein's equations cannot emerge since the classical condensate fluid equations are not fully background independent. In this sense, the lack of a well defined classical gravitational action on the bulk seems to prevent any advancement along this line of research.

Nonetheless, it was recently shown in Ref.~\cite{Belenchia:2014hga} that the field equations for N\"ordstrom gravity may emerge taking into account relativistic BECs in the massless limit. For such a realisation, it was assumed a real ground state of the aforementioned condensed matter system (constant phase) which implied a speed of sound equal to the speed of light which implies the absence, in this tuned system, of the Lorentz breaking transition between a low energy (phononic) Lorentz group and an high energy (atomic) one. Indeed, the main goal of Ref.~\cite{Belenchia:2014hga} consisted of showing that there can be an exact (not accidental) relativistic emergent gravity dynamics arising from an analogue system endowed with a UV completion provided by the atomic structure. 

Quite surprisingly, this investigations also showed that an unavoidable cosmological constant is induced in the emergent gravitational dynamics as a by product of the atomic interaction and the quantum non-equivalence of the phononic and atomic vacua. While the first contribution is positive for repulsive interactions, it was shown in \cite{Girelli:2008gc} that the second is generally negative (being related to the back-reaction of the atoms not in the condensate phase associated to the so called``depletion"). Then the  interplay between these two different contributions (which can be tuned to be of comparable size) can be used to mimic both positive and negative cosmological constants (and interestingly even transitions between opposite sign values).

Therefore, in such analogue systems one could mimic a well defined gravity theory in the bulk, i.e.~N\"ordstrom gravity, which, remarkably, can be described by an Einstein--Hilbert action with a cosmological constant plus a Lagrange multiplier term fixing the Weyl tensor to be zero~\cite{Deruelle:2011wu}. Indeed, this fact is quite relevant because it implies (see Appendix A for an explicit demonstration) that the boundary stress tensor for the bulk theory is still the standard Brown--York one~\cite{Brown:1992br} which in turns is known to take the form, for black hole geometries in AdS, of a perfect fluid stress energy tensor leading to the fluid/gravity duality. In this sense it would be quite interesting to be able to simulate AdS black holes-like geometries in this system.

In order to find such solutions, we first note that the required equality of the speed of sound to the speed of light implies that the acoustic disturbances in such special system propagates under an acoustic metric which is actually conformally flat (actually a necessary condition for them being solution of N\"ordstrom gravity) and given by 
\begin{equation}
g_{\mu\nu}=\phi^{2}\eta_{\mu\nu}, 
\end{equation}
where $\phi$ is the (real) ground state of the condensate. 

Indeed, there are static, spherically symmetric, and asymptotically flat solutions for N\"ordstrom gravity given by
\begin{equation}\label{NordstromBH}
ds^{2}=\left(1-\frac{m}{r}\right)\left[-dt^{2}+dr^{2}+r^{2}d\Omega^{2}\right]. 
\end{equation}
The above metric can be exactly mapped in both relativistic and non-relativistic BECs by imposing a fluid flow at rest and a density profile given by $\rho\approx\left(1-m/r\right)$. Albeit the above metric does not have the usual black hole properties, the existence of a Killing horizon is enough to ensure the existence of an acoustic black hole. 

Now, natural AdS generalisations of this solution could be provided e.g.~by taking the line elements 
\begin{eqnarray}\label{NordstromAdSBH}
ds^{2}&=&\left(1-\frac{m}{r}+\frac{r^2}{L^2}\right)\left[-dt^{2}+dr^{2}+r^{2}d\Omega^{2}\right],\\
ds^{2}&=&\left(1-\frac{m}{r}\right)\left(1+\frac{r^2}{L^2}\right)\left[-dt^{2}+dr^{2}+r^{2}d\Omega^{2}\right]. 
\end{eqnarray}
Unfortunately, none of the above (or similarly constructed) solutions seems to be able to satisfy the background fluid EOMs (which also coincide with the Einstein--Fokker equations of N\"ordstrom gravity with a cosmological constant). 

It is not clear if such an obstruction has a deep meaning about the limits of analogue gravity as an emergent gravity toy model (for example related to the issue raised in Ref.~\cite{Marolf:2014yga} about the notion of locality) or simply a technical issue which e.g. could be solved working perturbatively close to the N\"ordstrom limit of the relativistic BEC but still allowing for different speed of sound and light and hence full disformal geometries. All we can say for certain is that no no-go theorem seems to forbid AdS black holes solutions in N\"ordstrom gravity and that finding such solutions could pave the way to understanding the gravity/fluid duality at a fundamental level within a well-defined emergent gravity scenarios. For example, having such a solution would allow for a direct confrontation of the fluid dynamics induced on the boundary, via the computation of the renormalised stress energy tensor, with the underlying hydrodynamics in the bulk from which the structure of spacetime emerged. By doing so one could understand the origin of the fluid/gravity duality within an analogue setup (one way would be to obtain the transport coefficients of the induced boundary fluid system starting from the underlying fundamental fluid) and have a better understanding of it within an emergent gravity setup. We hope this investigation will stimulate further studies in this direction.

\acknowledgments

The authors wish to thank Sabine Hossenfelder and Thobias Zingg for useful discussions and remarks on the manuscript. Rodrigo Turcati is very grateful to CNPq for financial support. This publication was made possible through the support of a grant from the John Templeton Foundation, grant \# 51876. 
\appendix 

\section{Emergent Nordstrom gravity}
\label{app:1}

In the context of analoge gravity one might ask if the equations governing the dynamics of the quasi-particles (linearized fluctuations on the top of the relativistic condensate in this case) have  similarity 
with any known gravitational field equations. A striking resemblance was found in Ref.~\cite{Belenchia:2014hga} where it was shown that the fluctations of the condensate in the zero mass limit would experience a curved geometry determined by the  the Einstein--Fokker equation describing Nordstrom gravity. Using the Ricci scalar for the acoustic metric, which is given as $R=-6\frac{\square \phi}{\phi^3}$,  the relativistic Gross--Pitaevskii equation can be written as 
\begin{eqnarray} 
R+6\frac{m^2}{\phi^2}+12\lambda=\langle T \rangle,
\end{eqnarray}
where $ \langle T \rangle=-12\lambda [3\langle \psi_1 \rangle+\langle \psi_2 \rangle]$ and $\psi_1, \psi_2$ are the real and imaginary part of the fluctuation field. In the massless limit and after identifying the Newton and cosmological constants one gets the Einstein--Fokker equation 
\begin{eqnarray}
R-\Lambda=24 \pi \frac{G}{c^4}T.       
\end{eqnarray}
In the above equation we have not taken into account the quantum correctios to the cosmological constant but after considering it the sign of the cosmological constant might be different from what appears here. Remarkably this equation of motion can be derived from the Einstein--Hilbert action supplemented by a Lagrange multiplier. 

One can start with a covariant action given as
\begin{eqnarray} \label{action}
S[g_{\mu \nu}, \lambda_{\mu}{}^{\nu \rho \sigma}]=-\frac{c^3}{48\pi G}\int \sqrt{-g}\,d^dx \Big[R+\Lambda
\nonumber\\
+ \lambda_{\mu}{}^{\nu \rho \sigma}C^{\mu}{}_{\nu \rho \sigma}\Big],
\end{eqnarray}
where $G$ is the Newton constant, $R$ is the scalar curvature, $\Lambda$ is the cosmological constant,  $C^{\mu}{}_{\nu \rho \sigma}$ is the Weyl tensor and $ \lambda_{\mu}{}^{\nu \rho \sigma}$ is a Lagrange multiplier having all the symmetries of the Weyl tensor. Variation with respect to $g^{\mu \nu}$, imposing certain boundary condition on the variation of the metric and its first derivative and taking the trace part gives the Einstein–Fokker equation describing Nordstrom gravity with a negative cosmological constant, which is
\begin{eqnarray}
R-\Lambda=24 \pi \frac{G}{c^4}T,         \label{Neom}
\end{eqnarray}
and variation with respect to $\lambda_{\mu}{}^{\nu \rho \sigma}$ yields 
\begin{eqnarray}
C^{\mu}{}_{\nu \rho \sigma}=0,
\end{eqnarray}
which implies that the traceless part of the curvature is zero and we are left with only the trace part as required. Instead of imposing conditions on the first derivative of the metric we can add a counter term to the action whose variation would precisely cancel the variation of first derivative of metric term as obtained by extremization of the original action. The variation of the above action with respect to $g_{\mu \nu}$ can be written as
\begin{eqnarray}
\delta S[g_{\mu \nu}, \lambda_{\mu}{}^{\nu \rho \sigma}]=\frac{c^3}{48\pi G}\int d^4x \Big[-E^{\mu \nu}\delta g_{\mu \nu}+\partial_{\mu}V^{\mu}\Big] \nonumber\\
\end{eqnarray}
where $E^{\mu \nu}$ gives the field equation and the boundary term is given as
\begin{eqnarray}
V^{\mu}=-2D_{\sigma}\lambda ^{\mu (\nu \rho)\sigma} \delta g_{\nu \rho} +[g^{\nu \rho}g^{\mu}_{ \sigma}-g^{\mu(}g^{\rho)}{}_{\sigma}
\nonumber\\
-2\lambda^{\mu (\nu \rho)}{}_{\sigma}] \delta \Gamma^{\sigma}{}_{\nu \rho}.
\end{eqnarray}

In the above expression we can set the first term to zero by assuming the variation of the metric vanishes at the boundary. The second and the third term are canceled by adding the Gibbons hawking term to the the original action and the last term is the non zero part for the action we started with. Now we must note that all solution of the Nordstorm equation of motion must be conformally flat and for such metrics $ \delta \Gamma^{\sigma}{}_{\nu \rho} \sim B^{\sigma}\delta \eta_{\nu \rho} +\delta C^{\sigma}\eta_{\nu\rho}$. Here again the first term can be set to zero by our previous argument and the second term contacted with $\lambda^{\mu (\nu \rho)}{}_{\sigma}$ becomes zero due to the symmetries of $\lambda $. Hence the action (\ref{action}) with the counter terms becomes
\begin{eqnarray}   \label{nordstrom actionK}
S[g_{\mu \nu}, \lambda_{\mu}{}^{\nu \rho \sigma}]=-\frac{c^3}{48\pi G}\int_{\mathcal{M}} d^dx\sqrt{-g} \Big[R+\Lambda+ \lambda_{\mu}{}^{\nu \rho \sigma}C^{\mu}{}_{\nu \rho \sigma}\Big] 
\nonumber\\
-\frac{c^3}{24\pi G}\int_{\partial \mathcal{M}}d^{d-1}x \sqrt{-\gamma}K,
\nonumber\\
\end{eqnarray}
where $K$ is the trace of the extrinsic curvature and $\gamma$ is the determinant of the boundary metric $\gamma_{\alpha \beta}$. From the above action we can see that the boundary stress tensor of spacetime, i.e., the so-called Brown-York stress energy tensor, will have the same form as obtained for general relativity as it depends on the on-shell variation of the surface term with respect to the boundary metric and in this case we obtain the same boundary term as in the case of general relativity.

    \end{document}